\documentclass[a4paper,preprint]{revtex4}
\usepackage{graphicx}
\usepackage{dcolumn}
\usepackage{bm}

\begin{document}

\title{Transient electric fields in laser plasmas observed by proton streak deflectometry}

\author{T. Sokollik}
\email[]{Sokollik@mbi-berlin.de}
\author{M. Schn\"urer}
\author{S. Ter-Avetisyan}
\author{P.V. Nickles}
\author{E.Risse}
\author{M. Kalashnikov}
\author{W. Sandner}
\affiliation{Max Born Institut, Max Born Str. 2a, D-12489 Berlin}
\author{G. Priebe}
\affiliation{STFC Daresbury Laboratory, Warrington, Cheshire,WA4
4AD, Great Britain}
\author{M. Amin}
\author{T. Toncian}
\author{O. Willi}
\affiliation{Heinrich Heine Universit\"at D\"usseldorf, D-40225
D\"usseldorf, Germany}
\author{A.A. Andreev}
\affiliation{Vavilov State Optical Institute, St. Petersburg,
Russia}

\begin{abstract}
A novel proton imaging technique was applied which allows a
continuous temporal record of electric fields within a time window
of several nanoseconds. This "proton streak deflectometry" was
used to investigate transient electric fields of intense ($\sim
10^{17}$ W/cm$^{2}$) laser irradiated foils. We found out that
these fields with an absolute peak of up to $10^8$ V/m extend over
millimeter lateral extension and decay at nanosecond duration.
Hence, they last much longer than the ($\sim$ ps) laser
excitation, and extend much beyond the laser irradiation focus.\\
\\
\small{\textit{The following article has been submitted to Applied
Physics Letters. After it is published, it will be found at
http://apl.aip.org/.}}
\end{abstract}

\maketitle Laser proton acceleration is a rapidly emerging field
with yet unknown potential for beam specifications and
applications. The typical properties of the generated proton and
ion beams are picosecond emission, low longitudinal and
transversal emittance and in many cases a broad and continuous
kinetic energy distribution
\cite{Hatchett2000PoP,Borghesi2004PRL,Borghesi2003APL,Romagnani2005PRL,Toncian2006Science}.
These properties are almost exclusively determined by the
geoemtrical and temporal structure of the accelerating electric
field, created by the laser accelerated electrons. In this context
the recently observed phenomenon of transient electric fields with
large lateral extension (of the order of millimeters) has gained
considerable attention \cite{Toncian2006Science}. Such extension
exceeds the initial laser spot size considerably, hence, the
phenomenon of lateral electron transport \cite{McKenna2007PRL}
needs to be investigated for the complete understanding of
transient field geometries.

We employ a modified and extended version of the recently
developed method of proton beam deflectometry
\cite{Borghesi2003APL} and demonstrate this in a similar
experiment to \cite{Romagnani2005PRL} with lower laser irradiance.
If the proton beam is detected with a velocity dispersive detector
(e.g. a magnetic spectrometer) the influence of the transient
fields on the protons can be traced back in time because protons
with a specific energy arrive at the object to probe at a definite
time. Thus the velocity dependent detection of such a proton beam
can be named as "proton streak deflectometry". The advantage of
this method is the possibility of continuous recording of
transient fields on a ps time scale. In the presented experiments
the selected time window was about several nanoseconds with a time
resolution of about 30 ps.

\begin{figure}
\includegraphics{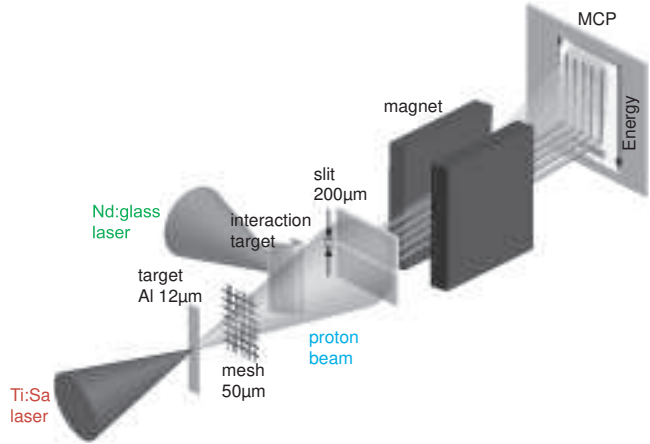}\caption{\small Experimental set-up for proton streak measurements. The
proton beam emerges from the rear side of a Ti:Sa laser irradiated foil. The proton beam is deflected by laser triggered fields from
the interaction target and analyzed due to the definite velocity dispersion of the magnet.}\label{fig1}
\end{figure}

In this work a proton beam streak technique is applied for the
first time to investigate the electric fields occurring at the
rear side of a laser irradiated thin metal foil. Two synchronized
high intensity laser at the Max-Born-Institute were employed for
the experiment. A high intensity ($\sim 10^{19}$ W/cm$^{2}$) 40 fs
Ti:Sapphire laser (CPA1) was used to produce the proton beam by
irradiating aluminium foils targets of 12 $\mu$m thickness which
are naturally covered with a water and hydro-carbon contamination
layer serving as a proton source. The second laser, a Nd:glass
laser (CPA2), was used to produce a second plasma which is to be
probed. This laser provides 1.5 ps laser pulses at a peak power of
about 5 TW and is synchronized to the CPA1 laser with an accuracy
of about 3 ps. The set-up of the experiment and the spectrometer
are depicted in Fig. \ref{fig1}.

The interaction target, a curved aluminium foil of 12 $\mu$m
thickness, was mounted as a stripe of about 8-10 mm width and bent
with a radius of about 5 mm. The CPA2  laser irradiated the
concave side and the proton beam probed the rear of the target at
90$^{\circ}$ to the target normal at the CPA2 interaction point.
The axis of the interaction target was at a distance of 40 mm from
the proton beam source. A mesh with a spacing of 50.8 $\mu$m
intersected the beam at a distance of 30 mm from the source. Two
imaging set-ups were employed: one with an entrance slit of 200
$\mu$m width to the magnetic spectrometer (set-up 1) and one in
which the entrance slit and the magnet were removed (set-up 2).
The slit to the spectrometer was placed on-axis to the Nd:glass
laser interaction point with an accuracy of $\pm\, 500 \, \mu m$.
The magnification of the interaction area was 16 fold in set-up 1
and 14 fold in set-up 2. The MCP detector \cite{Schreiber2006Pop}
and the phosphorous screen were gated in time in order to select
protons with a suitable time of flight (e.g. an applied 4 ns
gating selected protons with (1.4 - 2) MeV). The gating allowed us
to take snapshots in a way that is similar to the detection with
film stacks \cite{Borghesi2003APL, Romagnani2005PRL} where protons
of the same energy are all stopped in the same layer which is
almost exclusively exposed by these protons.

\begin{figure}
\includegraphics{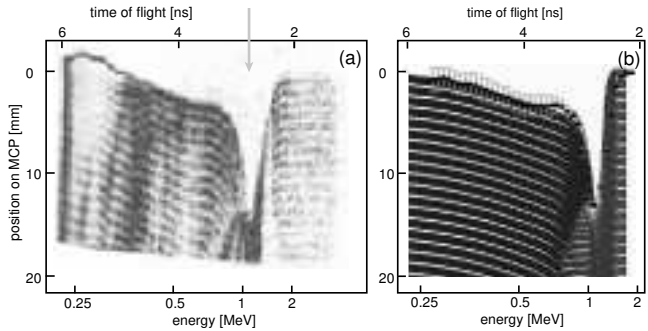}
\caption{Proton streak images and analysis: (a) - 2.3 J ($\sim 4 \cdot10^{17}$ W/cm$^{2}$)
pulse on the interaction target (also indicated in time by the arrow)
(b) - Simulated proton streak measurement including an extracted
deflection curve from the experimental data.}  \label{fig2}
\end{figure}

The transient field is generated at the rear side of a
laser-irradiated foil and hence the proton deflection relates to
the field strength which depends on the intensity of the incident
laser pulse. Along the x-axis the protons are dispersed according
to the velocity (energy) which determines the arrival time of the
probe pulse at the interaction target. In order to trace the
proton deflection perpendicular to the dispersion direction the
proton beam is intersected by a mesh. Each beamlet corresponds to
a small part of the proton beam which passes the interaction
target at a defined distance and its deflection can be projected
to the detector. This is possible due to the transverse laminarity
of our proton beam having an emittance value of 5 $\cdot10^{-3}
\pi$ mm mrad (method cf. \cite{Borghesi2004PRL}) for proton energy
above 1 MeV. Fig. \ref{fig2} (a)
 shows the "streaked" deflection of the proton
traces when the CPA2 pulse irradiates the interaction target at
intensities of $\sim 4\cdot10^{17}$ W/cm$^{2}$ (2.3 J pulse
energy). In Fig. \ref{fig2} (a) the sharp high energy cut-off of
the protons sets the maximum energy and the minimum is set by the
MCP gating. Both lasers are synchronized such, that the growth and
the decay of the deflecting field can be observed.

We analyzed our measurement with an analytical model. The model is
based on an empirical construction of two electrical fields which
change in space and time in order to explain the measured proton
beam deflection. As to be discussed in the following the
parameters suggest that these two fields can be associated with an
accelerated ion front \cite{Romagnani2005PRL} and a background
charge \cite{Beg2004}. Intensity and energy of the laser are used
to deduce an energetic electron population which propagates
through the target, spreads on the rear side and accounts for the
field generation. The proton ray-tracing is calculated in 3D
geometry and the final result is presented in Fig.2 B.
Additionally to the simulation result extracted deflection data
are inserted. The size of the error bars is caused by shot to shot
fluctuations of the proton beam pointing \cite{Schreiber2006Pop}
which occur at energies below 0.8 MeV. The variation of the
pointing has been calculated from 10 shots using CPA1 only.

As the experimental results suggest we calculate the deflection
with electric fields directed along the y-axis (cf. Fig.
\ref{fig1}). Without the entrance slit and the magnet (set-up 2),
two-dimensional snapshots of the deflected proton beam show that
the proton beam is diverted along the y- axis because the
grid-lines along y almost appear undistorted whereas the
grid-lines along the x-axis are shifted and compressed along y and
thus no longer visible. Furthermore we rely on the assumption that
magnetic fields which accompany an electron sheath are in first
order circular, with a symmetry axis parallel to the target
normal. Hence the action to the proton beam, traversing the
magnetic field perpendicular to its axis in a symmetrical ideal
case can be neglected.

The strong field component (Field 1) - a field front which decays
while propagating - causes a strong deflection of protons passing
the 2$^{nd}$ target near by and a decreasing deflection of protons
passing the 2$^{nd}$ target at larger distance. This feature is
visible at the rising edges and the maximum deflection (peak) of
the proton traces in the experiment (cf. Fig. \ref{fig2} (a)) and
in the model traces in Fig. \ref{fig2} (b). The weaker field
component (Field 2) - a coulomb field produced by the charge on
the cylinder surface - shifts the proton traces regardless to
their incoming original position on y-axis by a similar value
\cite{gerthsen1995}. This is visible at the lower energetic part
of the experimental traces as shown in Fig. 2 A and in the model
traces in Fig. \ref{fig2} (b). The combination of both describes
the whole recorded experimental picture. Taking the fluctuations
into account the determination of the strength of Field 1 and
Field 2 is subjected to an error of about 7 $\%$ and 20 $\%$,
respectively. The identified properties of Field 1 are similar to
those fields which occur during an ion front expansion. The 2D
grid picture supports the occurrence of two different fields: In
the color coded picture the yellow line is an area of an enhanced
proton number density caused by Field 1 while the shift and
blurring of the target edge can be attributed to Field 2.

Field 1 was constructed by modelling a one dimensional plasma
expansion into vacuum according to \cite{Romagnani2005PRL} and
\cite{Mora2003PRL}. Spatially, the electric field shows a plateau
region which is followed by an exponential rise up to the peak at
the front and then decays as ($1 + r / l)^{-1}$ where l is the
field scale length and r is the distance to the ion front. The
field front moves away from the target surface while the electric
field in the plateau region decays as ($1 + t / \tau)^{-2}$
whereas at the peak and the subsequent region the field decays as
$(1 + t / \tau)^{-1}$ (where $\tau$ is the decay time). Field 2 is
the field of a charged cylinder and it is supposed to be shielded
by the charge cloud accompanying the field front and thus to
influence only particles between the target surface and the front.
The electrons involved in the plasma expansion, assumed by the
scaling law given in \cite{fuchs} with a temperature of roughly
100 keV and carrying about 7.5 $\%$ of the focused laser energy
\cite{MS1995}, spread over the rear side of the target with a
Gaussian density distribution of about 6 mm FWHM. From the
experimental 2D spatially resolved pictures (cf. Fig. \ref{fig3})
it is visible that the distribution extends over several mm.
Correspondingly the simulation of the traces shows that smaller or
lager distributions can not account for the observed deflection
function. The field scale length (l) was supposed to be 100
$\mu$m, similar to \cite{Romagnani2005PRL}. The following
parameters could be also fitted to the experimental data: the
decay time $\tau$ (3 ps) of Field 1, the front propagation
velocity ($10^{6}$ m/s), the maximum charge density (Field 2) on
the target surface ($10^{-4}$ C/m$^{2}$), the linear grow within
10 ps and the exponential decay time (600 ps) of the target
charge. At t = 0 Field 1 dominates and peaks at the target surface
at about $3\cdot10^{8}$ V/m.

In summary we have demonstrated a novel imaging method, "proton
streak deflectometry", which allows measurements of the real-time
dynamics of transient intense fields in laser plasma interactions.
In particular we have investigated transient fields on the rear of
a laser irradiated metal foil which are responsible for the
process of laser proton acceleration. The observed streak images
were qualitatively explained by the temporal and 1D-spatial
development of two electric fields arising from charge-up and
charge compensation at a nanosecond timescale, and ion front
propagation at a timescale of several hundreds of picoseconds.
From that we conclude that we observed effects of energetic
electron generation and extended lateral transport which leads to
transient electric fields ($\sim 10^{8}$ V/m) with mm lateral
extension. Acknowledgement: This work was partly supported by DFG
- Sonderforschungsbereich Transregio TR18 and GRK 1203.
  \def\bibsection{\section*{}}%

\end{document}